\documentclass[prb,aps,super,
twocolumn,reprint,
groupedaddress,superscriptaddress,
amsfonts,amssymb,amsmath,floatfix,
showkeys,showpacs,a4paper]{revtex4-1}

\usepackage{graphicx}
\usepackage[english]{babel}
\usepackage{hyperref}
\usepackage{upgreek}
\usepackage{natbib}
\usepackage{lineno}
\usepackage{color}

\begin{document}

\title{Tailoring Magnetic Frustration in Strained Epitaxial  FeRh Films}

\author{Ralf Witte}\email{ralf.witte@kit.edu}
\affiliation{Institute of Nanotechnology, Karlsruhe Institute of Technology, 
 76344 Eggenstein-Leopoldshafen, Germany}
\author{Robert Kruk}
\affiliation{Institute of Nanotechnology, Karlsruhe Institute of Technology, 
 76344 Eggenstein-Leopoldshafen, Germany}
\author{Markus E. Gruner}
\affiliation{Faculty of Physics and Center for Nanointegration Duisburg-Essen (CENIDE), University of Duisburg-Essen, 47048 Duisburg, Germany}
\affiliation{Forschungs-Neutronenquelle Heinz Maier-Leibnitz (FRM II), Technische Universit\"at M\"unchen, 85748 Garching, Germany}
\author{Richard A. Brand}
\affiliation{Institute of Nanotechnology, Karlsruhe Institute of Technology, 
 76344 Eggenstein-Leopoldshafen, Germany}
\affiliation{Faculty of Physics and Center for Nanointegration Duisburg-Essen (CENIDE), University of Duisburg-Essen, 47048 Duisburg, Germany}
\author{Di Wang}
\affiliation{Institute of Nanotechnology, Karlsruhe Institute of Technology, 
 76344 Eggenstein-Leopoldshafen, Germany}
 \affiliation{Karlsruhe Nano Micro Facility (KNMF), Karlsruhe Institute of Technology, 76344 Eggenstein-Leopoldshafen, Germany}
\author{Sabine Schlabach}
\affiliation{Institute for Applied Materials, Karlsruhe Institute of Technology, 
 76344 Eggenstein-Leopoldshafen, Germany}
  \affiliation{Karlsruhe Nano Micro Facility (KNMF), Karlsruhe Institute of Technology, 76344 Eggenstein-Leopoldshafen, Germany}
\author{Andre Beck}
\affiliation{ Institute for Solid State Physics, Karlsruhe Institute of Technology, 
 76344 Eggenstein-Leopoldshafen, Germany}
\author{Virgil Provenzano}
\affiliation{Metallurgy Division, National Institute of Standards and Technology, Gaithersburg, Maryland 20899, USA}
\affiliation{Institute for Magnetics Research, George Washington University, Washington DC 20052, USA}
\author{ Rossitza Pentcheva}
\affiliation{Faculty of Physics and Center for Nanointegration Duisburg-Essen (CENIDE), University of Duisburg-Essen, 47048 Duisburg, Germany}
\author{Heiko Wende}
\affiliation{Faculty of Physics and Center for Nanointegration Duisburg-Essen (CENIDE), University of Duisburg-Essen, 47048 Duisburg, Germany}
\author{ Horst Hahn}
\affiliation{Institute of Nanotechnology, Karlsruhe Institute of Technology, 
 76344 Eggenstein-Leopoldshafen, Germany}
\affiliation{KIT-TUD-Joint Research Laboratory Nanomaterials, Technical University Darmstadt,  64287 Darmstadt, Germany}

\date{\today}
\begin{abstract}

We report on a strain-induced martensitic transformation, accompanied by a suppression of magnetic order in epitaxial films of chemically disordered FeRh. X-ray diffraction, transmission electron microscopy and electronic structure calculations reveal that the lowering of symmetry (from cubic to tetragonal) imposed by the epitaxial relation leads to a further, unexpected, tetragonal-to-orthorhombic transition, triggered by a band-Jahn-Teller-type lattice instability. 
 The collapse of magnetic order is a direct consequence of this structural change, which upsets the subtle balance between ferromagnetic nearest-neighbor interactions  arising from Fe-Rh hybridization and frustrated antiferromagnetic coupling among localized Fe moments at larger distances.

\end{abstract}


\maketitle

\section{Introduction}
  The discovery of the temperature-driven isostructural
  transition between  ferromagnetic (FM) and antiferromagnetic (AF) phases
  in chemically ordered (B2-phase, CsCl structure) equiatomic FeRh alloys nearly eight decades ago \cite{Fallot1938,Fallot1939}  posed a wide range of  intriguing fundamental problems and paved the way to possible applications. These comprise electric field control of magnetic order,\cite{Cherifi2014} heat assisted magnetic recording devices \cite{Thiele2003a} or a room temperature AF memory resistor. \cite{Marti2014,Moriyama2015}
  FeRh has also served as a model system for the investigation of magnetization reversal processes. \cite{Ju2004,Thiele2004}
  Last but not least, because of the large entropy change at the AF-FM transition, FeRh can be considered as one of the best magnetocaloric materials. \cite{Cooke2012,Liu2012,Stern-Taulats2015} Here we show that modifying this, already outstanding, material by applying expitaxial strain to a chemically \textit{disordered} metastable phase leads to a  hitherto unreported magnetostructural instability   in equiatomic FeRh.

Strain-control of material properties in general is a well-established concept in modern functional materials design. Epitaxial thin films can be considered as a special, technologically relevant, case.  Here the functional material is coherently grown onto a substrate and strain is applied by the mismatch between the two crystal lattices.
  To achieve extraordinarily strained films, it has been suggested to use soft materials \cite{VanDerMerwe1963} possessing inherent lattice instabilities, e.g materials undergoing martensitic transformations. \cite{Buschbeck2009}
Such experiments have been reported in epitaxially strained thin films of 
 disordered FePd and FePdCu  magnetic shape memory alloys (MSMA).    \cite{Buschbeck2009,  Kauffmann-Weiss2012b}  An alternative strategy for controlling the strain state in MSMA thin films is to use  ion-irradiation to create point defects which generate local stress fields in the material. \cite{Arabi-Hashemi2012}

 Concerning epitaxial B2-FeRh thin films a distinct  strain dependence of the AF-FM transition temperature  has been shown by several authors. \cite{Maat2005, Bordel2012, Cherifi2014,Loving2013,Lee2010_edit}  In contrast, little is known  about strain effects in the disordered alloys, where a significant impact  of interatomic distances \cite{Kudrnovsky2015} (i.e. strain) and chemical disorder \cite{Staunton2014} on the magnetic properties has been recently suggested by theory. 
 
  From an experimental point of view, the occurrence of martensitic lattice instabilities in the Fe-Rh system would be a crucial factor allowing for the growth of highly strained epitaxial films. 
 Indeed, such instabilities  have been reported in B2-ordered equiatomic  alloys induced by deformation \cite{Takahashi1995} or Pd-substitution \cite{Fukuda2013,Yuasa1995} and in disordered Fe-rich (20\,-\,25\,at\% Rh) alloys. Here, a transition from the face-centered-cubic (fcc) A1 phase to the body-centered-cubic (bcc) A2 phase was observed after quenching. \cite{Shirane1963,takahara1995} 
 
 The disordered bcc-A2 phase is suppressed in alloys with larger Rh content, due to the strong B2-ordering.\cite{Swartzendruber1984, Okamoto2011} 
   A thermodynamic analysis based on experimental data of different compositions and degree of order predicts a FM state with an estimated Curie temperature ($T_{\mathrm{C}}$) of 800\,K and  $\approx$4\,$\upmu_{\rm B}$ per unit cell  for A2-FeRh. \cite{Ohnuma2009} The long-range magnetic order of A2-FeRh arises from the dominating short-range FM exchange interactions between Fe-Fe and Fe-Rh pairs over competing AF  Fe-Fe interactions at larger separations \cite{Kudrnovsky2015,Ohnuma2009}  In turn, the net spin-polarization of the Rh orbitals is stabilized through the hybridization with the FM ordered Fe. \cite{Kudrnovsky2015,Sandratskii2011}

In the present study, we demonstrate that by carefully choosing the growth conditions an epitaxial strain-induced, adaptive martensite structure can be obtained in equiatomic, disordered FeRh thin films on W(001) (+6\% epitaxial mismatch). The initial lowering of symmetry (from cubic to tetragonal) 
induces a martensitic lattice instability which reduces the symmetry further towards a final orthorhombic structure, that grows with an adaptive nano-twinned martensite structure.  By taking advantage of the strong dependence of both the spin-polarization of the Rh orbitals and the Fe-Fe magnetic exchange parameters on the local structure, we effectively modify the balance between FM order and magnetic frustration and  possibly
achieve complete dynamic spin disorder  at low temperatures.

   This paper is organized as follows: In Sec.~\ref{sec:tech_exp} we provide
   a short description of the growth procedure of the heterostructure as well as details on the various applied characterization methods. Then we discuss in Sec.~\ref{sec:tech_dft} the technical details
   of the two complementary computational approaches used in this study.
   Sec.~\ref{sec:results_thinfilm} presents the strained epitaxial growth of FeRh on the W buffer-layer and it is shown by means of element-specific spectroscopy and magnetometry that  magnetic order is suppressed in the film. This peculiar experimental observation is investigated theoretically with super-cell calculations, presented in Sec.~\ref{sec:results_sc}, which reveal a cooperative  relaxation of the strained structure. This rearranged structure is then verified and refined experimentally by means of reciprocal space maps (RSMs in Sec.~\ref{sec:results_rsm}) and transmission electron microscopy (TEM in Sec.~\ref{sec:results_tem}). In Sec.~\ref{sec:results_kkr-cpa} we theoretically investigate the electronic structure and the magnetic properties  along a continuous transformation path, which finally explains the absence of magnetic order and identifies the underlying cause of the lattice instability. A summary and concluding remarks on the general implication of our findings are given in  Sec.~\ref{sec:sum}.

\section{Technical details}\label{sec:tech}
\subsection{Experimental details}\label{sec:tech_exp}

Epipolished MgO(001) single crystal substrates (5$\times$10$\times$0.5\,mm$^{3}$, \textit{SurfaceNET}) were  annealed and outgassed at 600\,$^{\circ}$\,C  in an oxygen atmosphere of 2.5$\times$10$^{-5}$\,mbar prior to deposition of the W layer. 
Afterwards the ~50\,nm W-buffer layer was deposited at 350\,$^{\circ}$C in a large-distance magnetron sputtering device \cite{Leufke2012} (base pressure of 2$\times$10$^{-9}$\,mbar) from a  W target (99.9\%, \textit{Haines und Maassen GmbH}). The growth rate used was ~0.022\,nm/s with an Ar backpressure of 0.0011\,mbar using a DC sputter power of 100\,W.
Equiatomic FeRh films ($\simeq $13\,nm in thickness, with isotopically enriched  $^{57}$Fe) were deposited at room temperature using a Mini e-beam evaporator (\textit{Oxford applied research}), which evaporates Rh from the rod (3\,mm diameter, 99.9\% \textit{Goodfellow}) and the $^{57}$Fe from a W crucible equipped with an alumina liner. The layer stack was then covered with a 1\,nm thick Rh capping layer for oxidation protection.
Epitaxial growth was monitored by reflection high energy electron diffraction (RHEED). 

 The structural details were studied by high resolution X-ray diffraction (HRXRD) with a \textit{Bruker} D8 Discover 4-circle diffractometer (Cu-K$_{\alpha}$ radiation), using a high-resolution setup with a Ge(022) 4-Bounce channel-cut monochromator. Rutherford backscattering spectroscopy confirms the composition to be Fe$_{52}$Rh$_{48}$ ($\pm$3\,at\%). The spectrum\cite{supplement} is measured with He ions at an incident energy of 2\,MeV, and it is fitted using the \textit{SIMNRA} 6.06 software. \cite{Mayer1999a}
 
A cross-sectional TEM specimen was cut from the film with a Focused-Ion-Beam (FIB) system (\textit{FEI} Strata 400 S). The initial cutting was performed using 30 kV Ga+ ions with final polishing at 5 kV and 2 kV. High-resolution transmission electron microscopy (HRTEM) was performed in a \textit{FEI} Titan 80-300 electron microscope, which was equipped with a CEOS image spherical aberration corrector and operated at an accelerating voltage of 300 kV.

The magnetic  properties of the thin film were characterized on a local  scale using $^{57}$Fe Conversion Electron M\"ossbauer Spectroscopy (CEMS). Low temperature CEMS was measured in a liquid nitrogen 
bath cryostat equipped with a proportional counter using He as the detector gas.  Magnetic characterization was performed using a MPMS XL \textit{Quantum Design} SQUID magnetometer.  The measured data were corrected for the diamagnetic contribution from the MgO substrate and a spurious magnetic signal coming from  paramagnetic impurities in the MgO \cite{Orna2010} (determined by measuring the pristine substrate): hence only the contribution from the FeRh/W film remains.

\subsection{Density functional theory calculations}\label{sec:tech_dft}
The interdependence of structural distortion and magnetic order is studied with two
complemetry approaches in the framework of density functional theory (DFT)
addressing the problem from the electronic structure point of view.
The first is based on super-cell calculations using the Vienna Ab-initio Simulation
Package (VASP).\cite{VASP1}
It describes the wavefunctions of the valence electrons
using a plane wave basis set while taking advantage of the
projector augmented wave (PAW) approach,\cite{VASP2}
which takes care of the
interaction with the core electrons. This usually yields
results comparable to all-electron methods.
In particular, we took the PAW-potentials from the standard database
generated for PBE-GGA (generalized
gradient approximation according to Perdew, Burke and
Ernzerhof)\cite{Perdew1996}
which consider explicitly the electronic configuration
$2p^63d^74s^1$ for Fe  and $4p^64d^85s^1$ for Rh (versions of Sep.\ 2000).
The cutoff energy of 366.5\,eV was chosen substantially above the cut-off requirement for the potentials
to guarantee accurately optimized structures.
For optimization, a $k$-mesh of 4$\times$4$\times$4 points was used together with the Brillouin zone
integration method of Methfessel and Paxton \cite{Methfessel1989} with a smearing of $\sigma=0.1\,$eV.
The final results were corroborated with a $k$-mesh of 6$\times$6$\times$6 points.
   In our VASP calculations, we modeled chemical disorder in the FM and PM phase using a
  32 atom special quasi-random structure (SQS), \cite{Zunger1990}
  developed by Jiang \cite{Jiang2009} for disordered ternary systems of the type A$_2$BC 
  (see Refs.~\cite{Jiang2009,supplement} for details).
  It thus allows for a disordered arrangement of collinear Fe-moments with opposite directions. 
The  local spin-polarization of the Rh $d$-electrons
  is determined by the hybridization with the surrounding Fe states. \cite{Sandratskii2011,Kudrnovsky2015}

The second approach is based on the
Korringa-Kohn-Rostoker (KKR) method as implemented in the Munich SPR-KKR
code.\cite{SPR-KKR1,SPR-KKR2,Ebert2011}
Angular momentum expansion was taken into account up to $l_{\rm max}=f$ and up to
64 points were used to describe energy contour in the complex plane
in combination with a large k-mesh (up to 49$\times$49$\times$26
in the full Brillouin zone, i.\,e., 8750 points in the irreducible zone).
A scalar relativistic description of the Hamiltonian
has been used in combination with the atomic sphere approximation
(ASA). The exchange-correlation functional was treated within the PBE-GGA.\cite{Perdew1996}
The chemical disorder was modeled analytically 
within the single-site coherent potential approximation (CPA), which allows for
the representation of perfectly disordered systems in small unit cells
but not for the individual relaxation of atomic positions.
For the magnetically disordered PM configuration, we used a simplified
disordered local moment approach by taking again  advantage of the CPA, which allowed
us to model a fractional occupation of the same lattice site with
Fe-atoms with oppositely oriented magnetic moments.
The Heisenberg model exchange parameters were determined using the approach of Liechtenstein
{\em et al.}\cite{Liechtenstein1987}

\section{Results and discussion}\label{sec:results}
\subsection{Thin film growth and characterization}\label{sec:results_thinfilm}
\begin{figure}[htb]
\[\includegraphics[width=0.8\columnwidth]{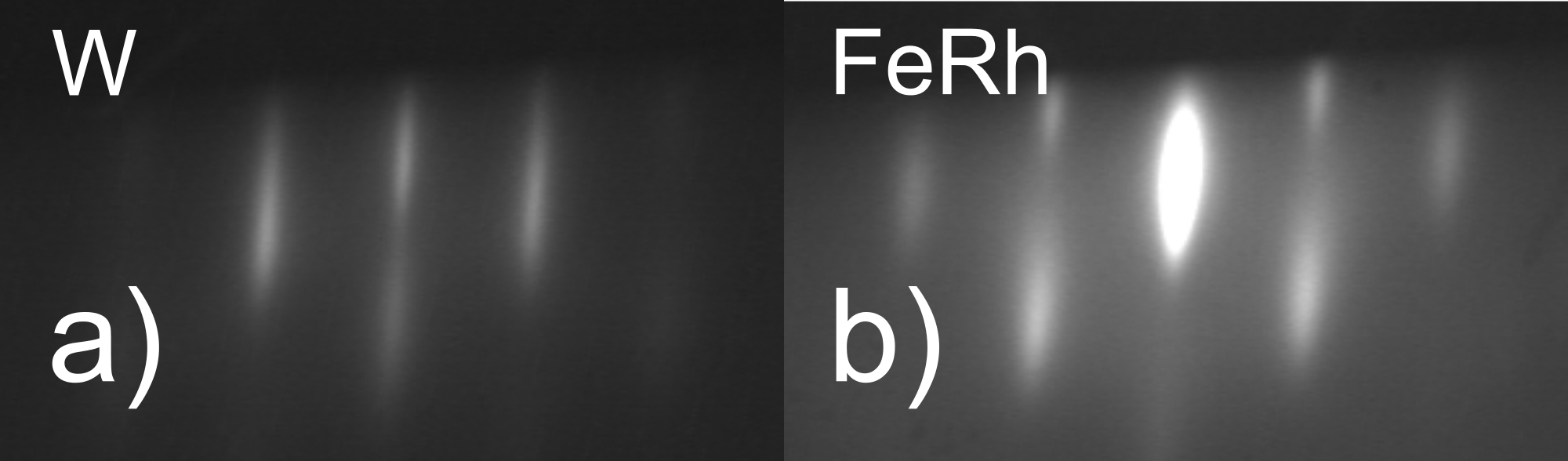}\]
\caption{(a) and (b): RHEED patterns for W(001) and FeRh surfaces respectively, the electron beam points in [010] direction of the bcc lattice.}
\label{figure1}
\end{figure}
  Fig.\,\ref{figure1}a)  presents the RHEED pattern along the [010] direction of the (001) W-bcc lattice: the pattern is sharp and streaky, typical for a flat, single crystalline surface. During the deposition of the FeRh  layer the pattern stays streaky  indicating again epitaxial growth as shown by the final pattern in Fig.~\ref{figure1}b).

\begin{figure}[htb]
\[\includegraphics[width=1\columnwidth]{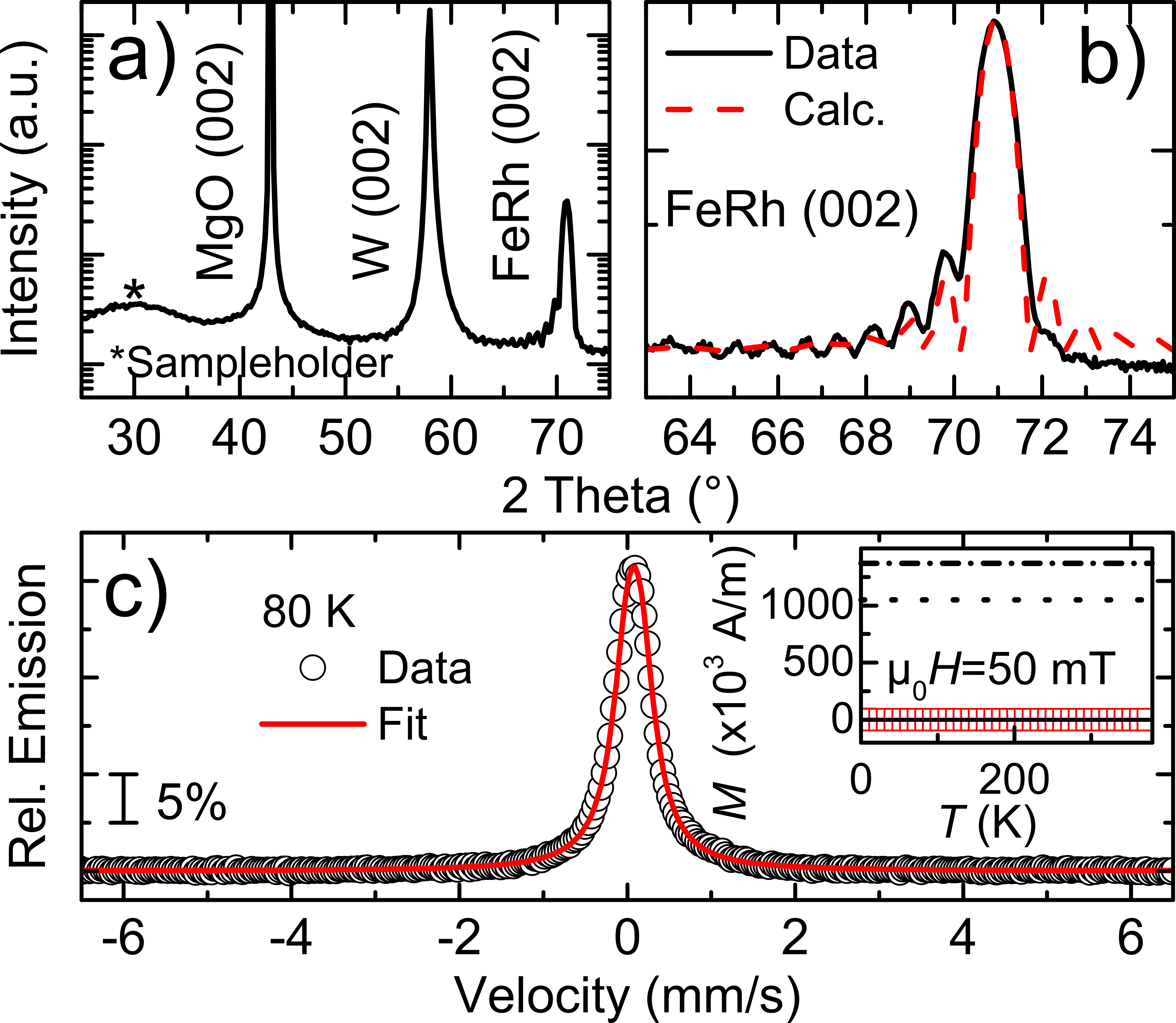}\]
\caption{(Color online) a) HRXRD pattern
 showing the (002) reflections for W and FeRh. b) Detailed view of (002) FeRh showing Laue oscillations. The data is reproduced using the Laue-equation. c) CEMS spectrum at 80\,K, showing no sign of magnetic splitting. The data have been fitted with a quadrupole doublet. Inset: SQUID, Field cooling curve $M(T)$ in a magnetic field $\upmu_{0}H$ of 50\,mT, $T=$5\,-\,380\,K. The two lines indicate magnetization values expected for A2 (dash-dotted) \cite{Ohnuma2009} or partially B2-ordered (dotted) \cite{Lu2013} FeRh.
}
\label{figure2}
\end{figure}

The diffraction pattern in Fig.~\ref{figure2}a) shows two reflections, in addition to the MgO(002) reflection,  attributed to the (002) planes of W and of FeRh, respectively. The out-of-plane spacing of the W is determined to be 3.179(1)\,\AA, which is close to the literature value. \cite{Chiarotti1993} Reciprocal space maps (RSMs) on asymmetric W reflections  (having an in-plane component of the scattering vector), yield  3.151(2)\,\AA\ in-plane lattice constant, suggesting an almost fully relaxed growth of the W buffer-layer on MgO(001). \cite{supplement}

The FeRh reflection reveals a lattice constant of 2.655(5)\,\AA\,, implying a contraction of -11\% compared to 2.989\,\AA\, for B2-FeRh in the AF state.\cite{Swartzendruber1984} The reflection  exhibits Laue oscillations (see Fig. \ref{figure2}b)), due to the finite thickness of the layer.  This proves fully coherent growth with an uniform lattice constant through the entire film,  excluding gradual strain relaxation. The out-of-plane contraction ($c/a\sim $\,0.84) is due to the large in-plane tensile, epitaxial strain (+6\%).  It is discernible that the FeRh alloy is in a disordered structure as manifested by the absence of the (001) superstructure reflection at $\sim $35$^\circ$.

The magnetic properties of the FeRh thin film were investigated with CEMS, which is sensitive to the local magnetic hyperfine field at the $^{57}$Fe nucleus sites and, as an element specific method, it measures the magnetic properties of Fe in the FeRh film only. In CEMS any static magnetic Fe phase, irrespective of its nature, FM, AF, non-collinear or spin glass, would be reflected in a characteristic six-line magnetic splitting. 
 The absence of this feature in the spectrum measured at 80\,K in Fig.~\ref{figure2}\,c)  shows that no magnetic order is present. Instead, the spectrum is fitted best with a broad quadrupole doublet.\cite{supplement}
   The inset in Fig.~\ref{figure2}\,c) displays a field-cooling curve $M(T)$\, of the sample measured in $\upmu_0H=50$\,mT in the SQUID. As a reference the saturation magnetization values expected for A2- \cite{Ohnuma2009}  or partial B2-FeRh \cite{Lu2013}  are also plotted. 
   
 Based on both, the featureless shape of the field-cooling curve and the CEMS results we can conclude that there is no  magnetically  ordered phase at least down to 5\,K. Hence, magnetic order is effectively suppressed in the highly strained FeRh film.  

\subsection{Super-cell calculations}\label{sec:results_sc}
The super-cell calculations with the VASP package, described in Sec.~\ref{sec:tech}, provided the
first step to unravel the link between structural distortion and magnetic order on the atomic scale.
In the beginning, we compared FM and PM configurations in a  body-centered-tetragonal (bct) environment with $c/a=0.84$ (experimental value), but with atoms fixed at the ideal lattice positions.  
  Here, FM order is favored  as in the cubic A2 case, while a downhill relaxation of the atomic positions within the fixed super-cell  leads to a PM ground state.

Since distribution of the atoms in the SQS cell does not obey cubic symmetry, we imposed the tetragonal distortion with the compressed $c$-axis oriented parallel to either of the three cartesian axes. This leads to essentially similar structures in terms of the relaxation pattern at numerically different energies. However, the qualitative results are deemed independent on
the specific decoration of the cell, since
we consistently found in all calculations the same energetic trend: A PM ground state in the relaxed structures, in contrast to the FM preferred in all unrelaxed tetragonal super-cells.

\begin{figure}[htb]
\[\includegraphics[width=\columnwidth]{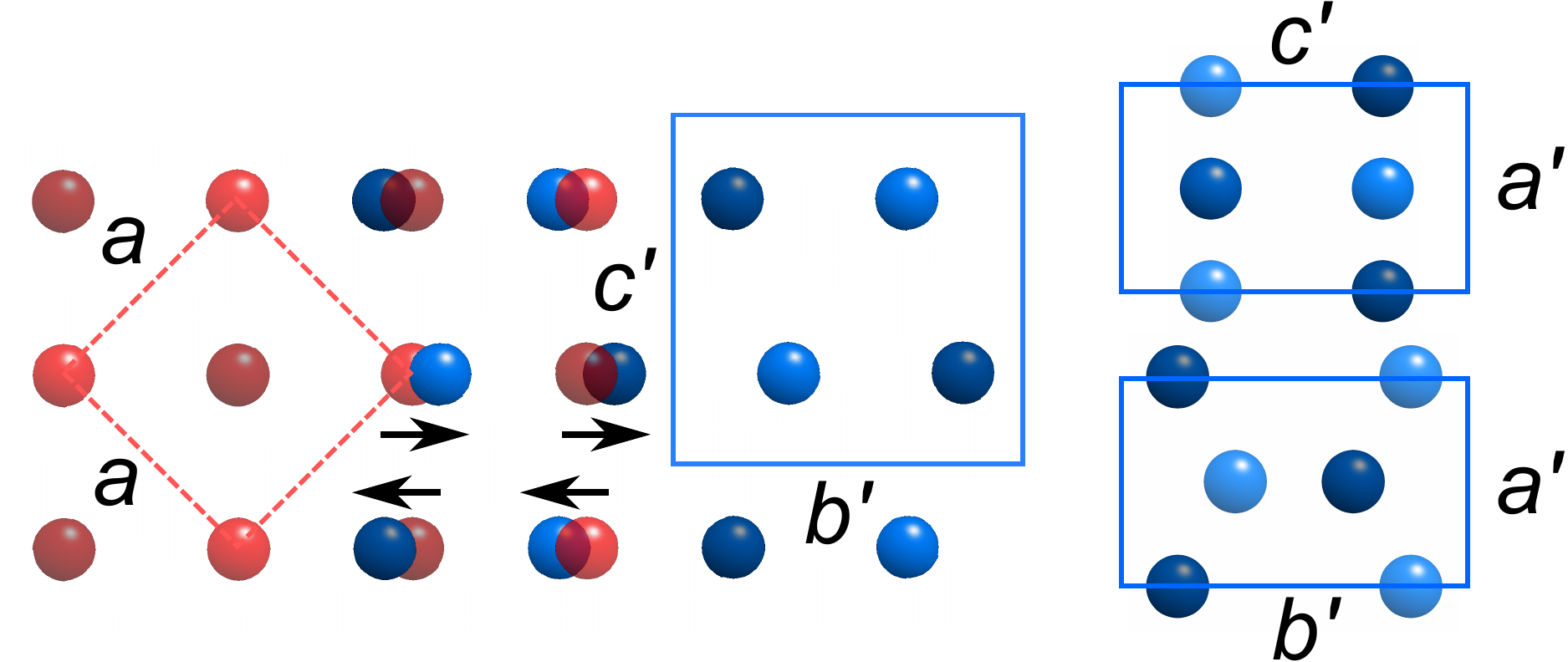}\]
\caption{(Color on-line) View of the (left, red) bct parental lattice and the orthorhombic structure obtained by the super-cell calculation (right, blue, along all three crystallographic directions), (middle) superposition of both structures illustrating the cooperative motion of the atoms (arrows).}
\label{figure3}
\end{figure}

   The relaxation causes a cooperative displacement of the atoms from their ideal bct sites,  breaking  tetragonal symmetry.   A geometrical analysis \cite{supplement} reveals that the relaxation  leads to a redefined unit cell with atomic positions, which can be described in the orthorhombic space group No. 63 (\textit{Cmcm}).\cite{Stokes2005} In the orthorhombic cell, the highest symmetry axis $a^{\prime}=c$ is the out-of-plane axis of the film, while $b^{\prime}=c^{\prime}=\sqrt{a}$ are the in-plane axes, rotated by 45$^{\circ}$ with respect to the primary tetragonal cell. The structure can be described with only one 4c Wyckoff position (0,$y$,0.25) with $y=-0.17$, which describes essentially a displacement of the atoms along the $b^{\prime}$-axis. The parental bct structure is recovered by setting $y=-0.25$.
The two main characteristics of the structural rearrangement (cooperative displacement and lowering of symmetry) underline the martensitic nature of the observed phenomenon.
Fig.~\ref{figure3} illustrates the rearrangement in a top-view of the atomic columns of the parental bct (left), the relaxed structure (right) and a superposition of both (middle).

\subsection{Reciprocal space maps}\label{sec:results_rsm}
\begin{figure}[htb]
\[\includegraphics[width=\columnwidth]{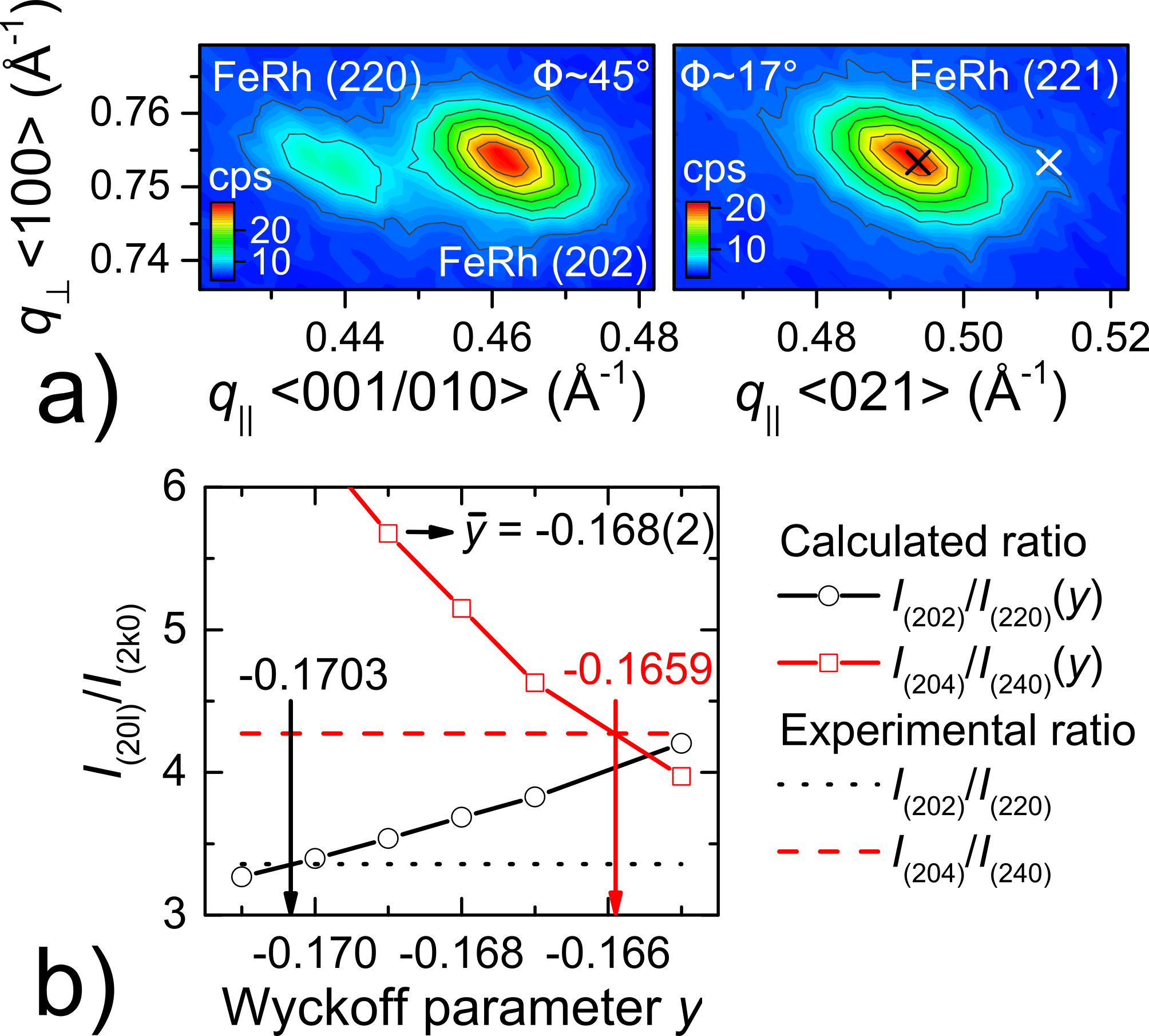}\]
\caption{(Color on-line)  a) RSM intensity plots of asymmetric reflections:  (202)/(220) showing the difference in in-plane lattice constants as well as their differing intensity, (221) proving the orthorhombic symmetry. The black and white cross depict the calculated position for (221) and a possible (212) reflection, respectively. See text for details. b) Plot of the calculated intensity ration of $ I_{(20l)}/I_{/2k0)}$ reflections as function of the $y$ parameter of the 4c Wyckoff site. The measured intensity ratio is plotted as straight line.}
\label{figure4}
\end{figure}

The orthorhombic structure identified in Sec.~\ref{sec:results_sc} must result in an additional set of Bragg reflections, compared to the bct structure, which can be investigated in RSMs. Fig.~\ref{figure4}\,a) presents two RSMs measured under  different azimuthal angles $\upphi$ (0$^{\circ}$ corresponds to the principal axes of the bcc-W lattice). The first one, displayed in the left panel, is recorded with $\upphi=45^{\circ}$, hence parallel to the orthorhombic lattice. It displays the part of reciprocal space where the  (202) and (220) reflections of the orthorhombic structure are expected.

The intensity plot depicts two clearly split reflections, which differ also in their intensity. The reflections
 are observed individually at different in-plane scattering vectors $q_{||}$, while their positions in the out-of-plane direction $q_{\perp}$ are basically identical. This is  evidence for non-equivalent in-plane lattice parameters,  which can be determined to be $b^{\prime}=4.584(14)$\,\AA, $c^{\prime}=4.324(6)$\,\AA. Therefore not only the change in $y$, predicted by theory, but also the in-plane dimensions of the unit cell reflect the orthorhombic distortion.

  Another important result is the simultaneous observation of the two reflections in the same azimuthal direction, which implies that the film must consist of 90$^{\circ}$ rotated variants (domains), similar to the adaptive  nano\-twinned  martensites in Ni$_{2}$MnGa \cite{Kaufmann2010a} and FePd \cite{Kauffmann-Weiss2011} thin films. At these two points the experimental results go beyond the theoretical predictions as these particular structural features were not accounted for in the super-cell calculations. It is noteworthy that the orthorhombic unit cell still matches the W lattice, when considering that $\sqrt{b^{\prime 2}+c^{\prime 2}}\approx 2a_{\rm W}$, underlining the adaptive nature of the rearranged structure.

An experimental approach to refine all crystallographic parameters independently, as in a full pattern fitting of integrated intensities for a single crystal, is very challenging in the case of epitaxial thin films. \cite{supplement} 
Hence, a  full pattern fitting is not feasible in the present case, but it will be shown that a quantitative refinement of the structure can still be achieved by combining the results of the super-cell calculation  with the experimental data.

As a starting model we use the structure (spacegroup No. 63, \textit{Cmcm}, 4c Wyckoff site with $y=-0.17$) obtained in super-cell modeling, with the above determined orthorhombic lattice parameters. Hence, the only structural parameter that requires refinement is the $y$ parameter of the 4c Wyckoff site (the others are fixed, due to symmetry requirements). The $y$ parameter, which essentially describes the displacement of the atoms along $b^{\prime}$, is the major factor governing the intensity difference between reflections having either ($ hk0 $) or ($h0l$) components. So by comparing the experimental ratio of integrated intensities of one pair $ I_{(h0l)}/I_{(hk0)}$ with calculated intensity ratios as function of $y$, we can quantitatively determine $y$.\cite{supplement} Hereby we make use of the fact, that the observation of both in-plane directions $b^{\prime}$ and $c^{\prime}$ in the same azimuthal direction and their proximity in reciprocal (and also angular) space allows us  to neglect, in first approximation, all scattering geometry dependent factors (since they cancel out under the division).

Figure \ref{figure4}b) depicts the calculated intensity ratios as function of $y$ and the experimentally obtained ratios (displayed as constant line) for the (202)/(220) pair as well as for the higher order reflection (204)/(204). From the intersection of the calculated curves with the corresponding experimental value a certain $y$ value is determined, which yields a mean value of   -0.168(2), in full agreement with the results of the super-cell calculations. The  crystallographic parameters are summarized in Tab.~\ref{table_crys_para}.  

\begin{table}[htb]
\centering
\caption{Crystallogrpahic parameters of the orthorhombic structure.}
\label{table_crys_para}
 \begin{ruledtabular}
\begin{tabular}{ccccc}

      \multicolumn{5}{c}{Spacegroup No. 63, \textit{Cmcm}} \\
   \hline
		\multicolumn{3}{l}{Lattice parameter (\AA)} & \multicolumn{2}{l}{Wyckoff site 4c} \\
            
        $a^{\prime}$ &  \multicolumn{2}{l}{2.655(5)} & \textit{x} & 0     \\
        $b^{\prime}$ & \multicolumn{2}{l}{4.584(14)} & \textit{y} &-0.168(2)    \\
         $c^{\prime}$ & \multicolumn{2}{l}{4.324(6)} & \textit{ z} & 0.25   \\

\end{tabular}
\end{ruledtabular}
\end{table}

Of course, as already pointed out above, the orthorhombic structure must result in a different set of Bragg reflections compared to the parental bct lattice. 19 out of 25 geometrically accessible reflections\footnote{The six remaining reflections were too weak and/or not observable due to an unfavorably large angle between respective plane and sample surface} could be measured in RSMs, 14 out of the observed reflections are only present for the orthorhombic structure.\cite{supplement} An exemplary orthorhombic reflection is displayed on the right-hand side of Fig.~\ref{figure4}a), namely the (221), which is observed for $\upphi\approx 17^{\circ}$. Its  position, calculated using the crystallographic parameters shown in Tab.~\ref{table_crys_para}, is indicated by a black cross. The fact that  only this very reflection appears here, and not also the (212)  reflections (in-plane indices swapped, its expected position is marked by the white cross), shows that the diffraction follows the general reflection conditions $hkl$,~$h+k=2n$ of the $Cmcm$ space-group.\cite{Aroyo2006} This observation  provides ultimate proof that the  structure model, derived in the combined experimental and theoretical approach, is in agreement with the diffraction data.

\subsection{Transmission electron microscopy}\label{sec:results_tem}
\begin{figure}[htb]
\[\includegraphics[width=\columnwidth]{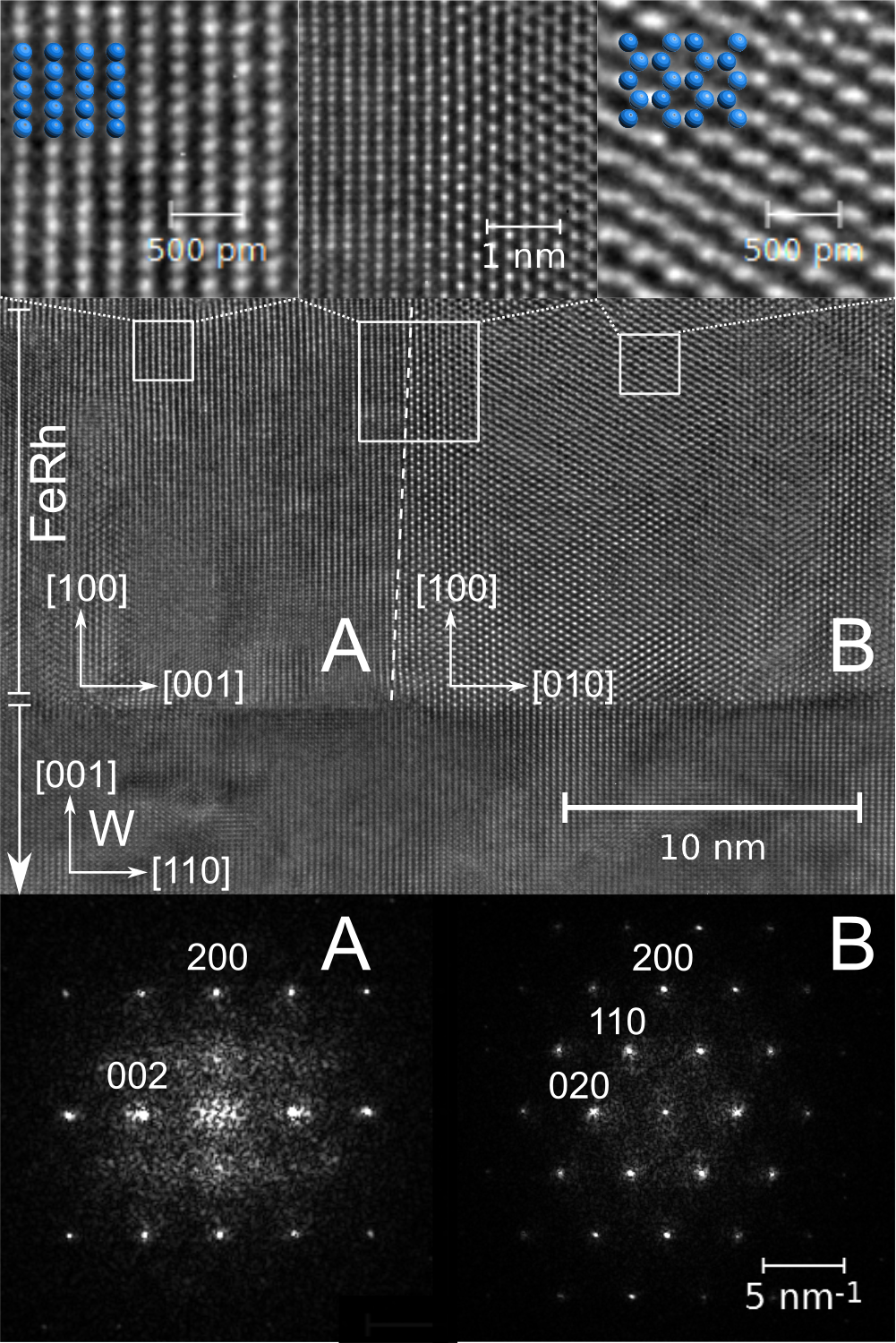}\]
\caption{(Color on-line) Results of TEM investigation. Center part: Cross-sectional view of the FeRh/W multilayer structure. In the FeRh two distinct domains with 90$^{\circ}$ rotated in-plane orientation are observed (separated by the dashed line). Bottom-row: respective FFT of the two structural domains. Top-row: Magnified images of the indicated areas, showing the projection of the structural model together with the experimental data.  The center image displays the interface between both domains. }
\label{figure6}
\end{figure}
A  local structural view is obtained by HRTEM. A cross-sectional view of  the film is shown in Fig.~\ref{figure6}. The FIB lamella was cut such that the electron beam is nearly parallel to the $<$110$>$ direction of the W-buffer layer and hence also to the orthorhombic in-plane axes of the FeRh lattice. Indeed, the center part of the figure shows a representative high-resolution micrograph featuring the top part of the W buffer-layer and two FeRh grains that are oriented 90$^{\circ}$  with respect to each other in the plane. The dashed line indicates the boundary between the grains.

 The Fast-Fourier-Transform (FFT) from each grain shown below allows to determine the epitaxial relation of the two grains. Grain A is oriented with FeRh$_{Cmcm}$[001]\,$\parallel$\,W[110] (horizontal direction in the figure), while for grain B the same direction corresponds to FeRh$_{Cmcm}$[010]. Both grains are coherently aligned with their (100) planes parallel to the FeRh/W interface.
 The  different orientations become  apparent when comparing the two FFT patterns. FFT of grain A shows no (101) reflection while the (110) in B is present, which is another indication for the above presented analysis of crystallographic symmetry as it fits to the systematic absence of space-group $Cmcm$  (reflection condition: $hk0$,~$h+k=2n$ and $h0l$,~$h,l=2n$).\cite{Aroyo2006} Furthermore, the magnified areas presented on top in direct  comparison with the atom-ball model, unambiguously illustrate the full-agreement between the proposed structural  model and the local lattice structure observed in the electron microscope.

The interface area between the two structural domains is magnified in the top-center image. It clearly depicts that both  grains are coherently oriented in [100] out of-plane direction.
 Moreover from the obtained images an average size of the domains of about 10\,-\,20\,nm can be determined. This  nanometer-scale structure supports that strain-adaption is one driving force for the structural re-arrangements.

\subsection{Magnetic and electronic structure}\label{sec:results_kkr-cpa}
 The observed lack of magnetic order in the disordered FeRh films on W provided the interest in investigating the details of the resulting crystallographic structure. This led to the
  realization that the tetragonal distortion induces a further structural relaxation and results in a final orthorhombic structure.  This is accounted for by the  \textit{Cmcm} primitive cell with the experimentally determined structural parameters $c/a$, $y$ and $c^{\prime}/b^{\prime}$. Hence, with this information, we are able to
  investigate  the electronic and magnetic properties in a small super-cell with the
  KKR-CPA approach. These calculations are carried out
  along a hypothetical transformation path from the cubic A2 structure
  to the final orthorhombic structure, as obtained from experiment.
  Along the path, we have kept the volume per atom constant.
    \begin{figure}[htb]
\[\includegraphics[width=\columnwidth]{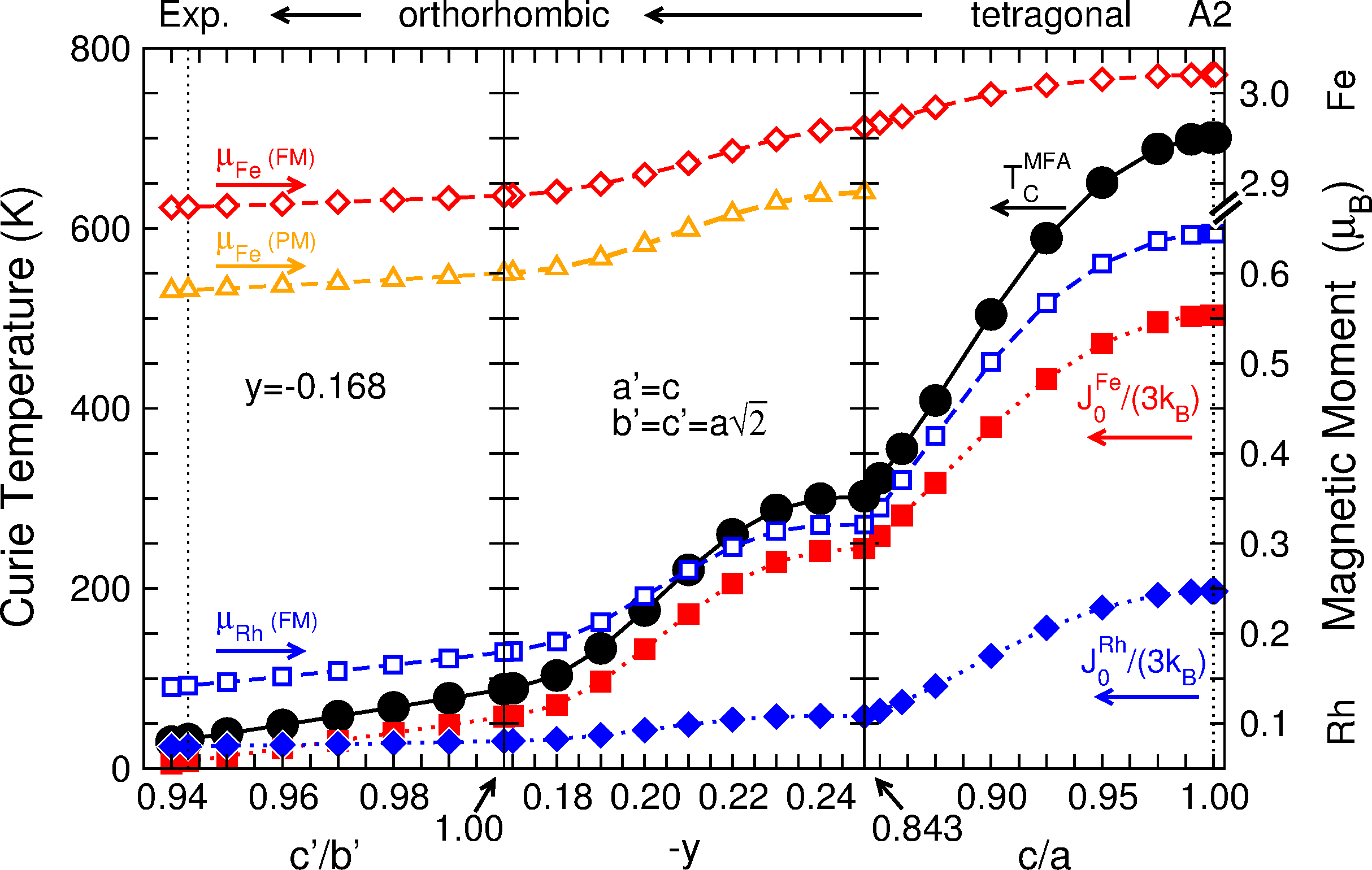}\]
\caption{
  (Color online) Calculated variation of effective exchange coupling constants $J_0^{\rm Fe,Rh}$,
  mean-field Curie temperature $T_{\rm C}^{\rm MFA}$, Rh and Fe magnetic moments, $\upmu_{\rm Rh}$ and $\upmu_{\rm Fe}$
  (the latter for the FM and the PM state) along
  a hypothetical transformation path, from right to left:
  bcc-A2$\rightarrow$bct\,($c/a$=$0.843$)\,$\rightarrow$\textit{Cmcm}\,($y$=$-0.168$,\,$b^{\prime}$=$c^{\prime}$)
  $\rightarrow$\textit{Cmcm}\,($y$=$-0.168$,\,$c^{\prime}/b^{\prime}$=$0.943$). 
}\label{figure7}
\end{figure}
         
\begin{figure}[htb]
\[\includegraphics[width=0.9\columnwidth]{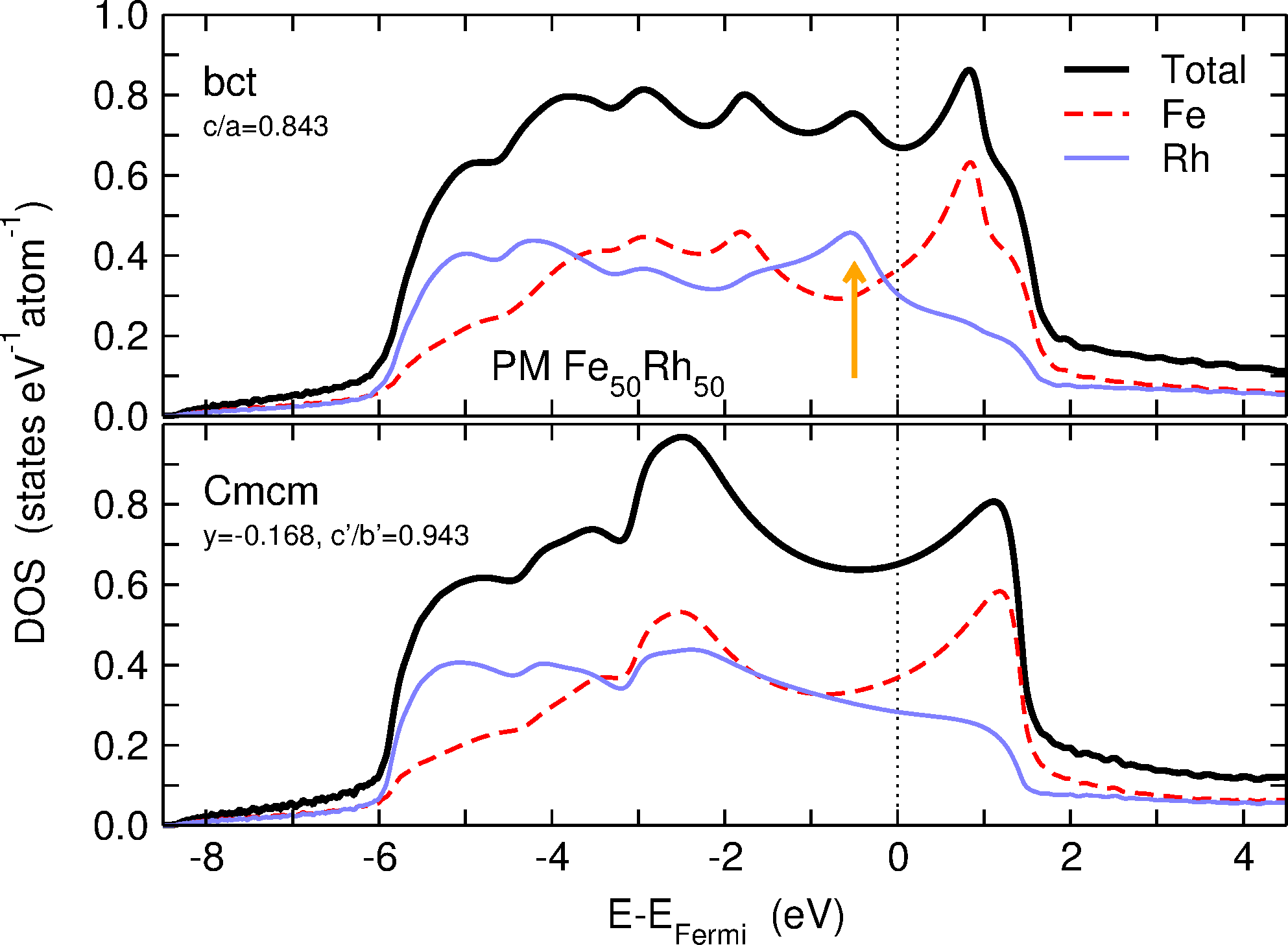}\]
\caption{
  (Color online) Element resolved electronic density of states (DOS) of PM FeRh in the tetragonal state with $c/a$\,=\,$0.843$ (top) and the orthorhombic ground state (bottom). The orange arrow indicates the  peak in the Rh-DOS, which is redistributed upon structural relaxation.
}\label{figure9}
\end{figure}
\begin{figure}[htb]
\[\includegraphics[width=0.9\columnwidth]{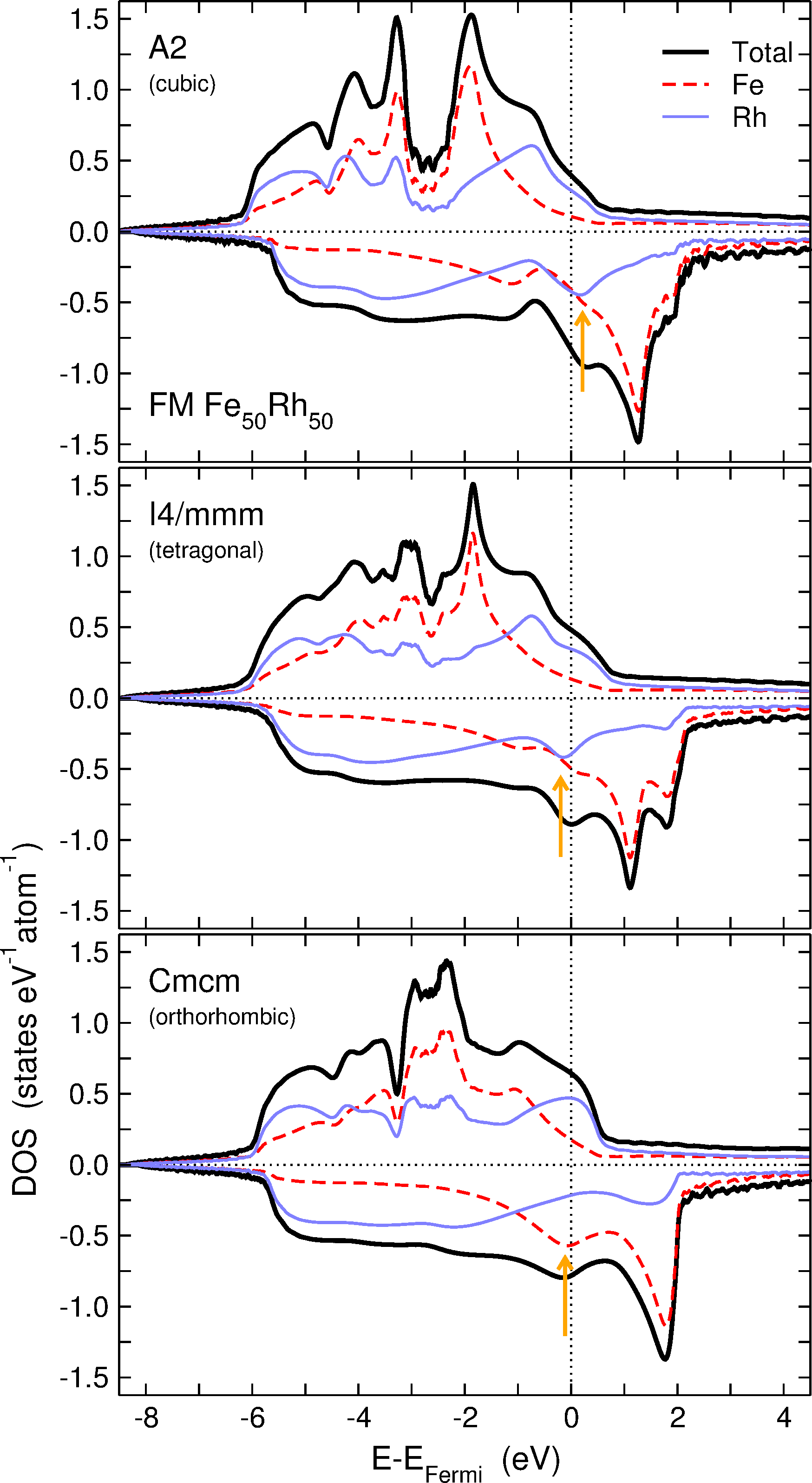}\]
\caption{Total and element resolved electronic DOS of cubic FM A2-FeRh (top), tetragonal (center) and orthorhombic
(bottom) FeRh. The features discussed in the text are marked by orange arrows.}
\label{figure10}
\end{figure}
\begin{figure*}[htb]
\[\includegraphics[width=0.8\textwidth]{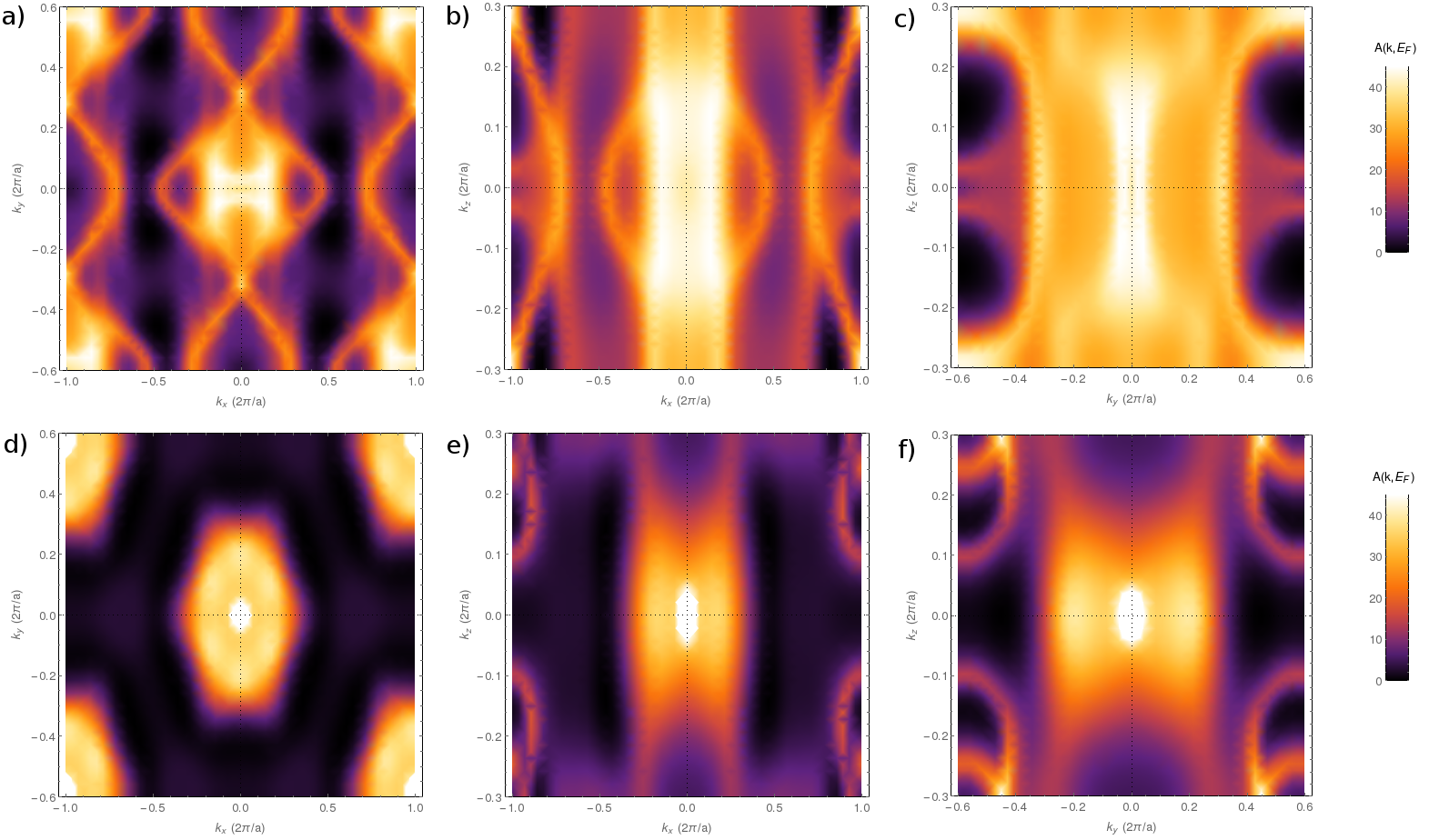}\]
\caption{(Color online)Cross-sections through the minority spin Fermi surface of FM FeRh, in terms of the Bloch spectral
  function $A(\vec{k},E_{\rm F})$ represented in the reciprocal coordinates of the orthorhombic lattice.
  The subfigures (a)-(c) refer to the $k_x$-$k_y$, the $k_x$-$k_z$ and the $k_y$-$k_z$ plane of the structure with tetragonal symmetry,
  while (d)-(e) show the corresponding plots of the final structure with orthorhombic symmetry.
  The colors on the intensity scale indicate to the magnitude of $A(\vec{k},E_{\rm F})$
  in arbitrary units.}
\label{figure11}
\end{figure*}

The energetic order of the magnetic and geometric structures
and the respective changes in the electronic density-of-states (DOS)
obtained with the KKR-CPA approach
were found to be consistent with the super-cell calculations described above.
The $T_{\mathrm{C}}$ in the mean-field approximation (MFA) is directly linked to the 
  element resolved effective intersite exchange parameters $J_0^{\rm Fe}$ and $J_0^{\rm Rh}$ by
  $k_{\rm B}T_{\rm C}^{\rm MFA}$$\,=\,$$\frac{2}{3}\,J_0$$\,=\,$$\frac{2}{3}\,(0.5\,J_0^{\rm Fe}+0.5\,J_0^{\rm Rh})$ calculated using the approach of Liechtenstein and co-workers. \cite{Liechtenstein1987}
  
  The variation of $T_{\rm C}^{\rm MFA}$, its element-resolved constituents and the magnetic moments
  $\upmu_{\rm Fe}$, $\upmu_{\rm Rh}$ are depicted in Fig.~\ref{figure7}.
   Starting from the right-hand side with the cubic A2 structure, we first tetragonally distort the cell, before lowering the symmetry  to orthorhombic by varying the internal $y$-parameter.
  Finally, the ratio $c^{\prime}/b^{\prime}$ is altered without changing the symmetry any further.
    With increasing tetragonal distortion, $T_{\rm C}^{\rm MFA}$
  decreases rapidly,  due to a reduction of both $J_0^{\rm Fe}$  and $J_0^{\rm Rh}$. The drastic decrease of the latter is directly linked to the altered hybridization between Fe and Rh states under these structural changes, leading to a reduced spin-polarization of Rh $d$-electrons and thus to a smaller $\upmu_{\rm Rh}$. 
  Since the contributions to $J_0^{\rm Rh}$ originate mostly from FM coupled Fe-Rh pairs,\cite{Sandratskii2011,Kudrnovsky2015} the stronger  decrease in $J_0^{\rm Fe}$ cannot be explained by the reduced FM Fe-Rh interaction alone. This indicates an increasing competition between the present short-range FM and longer ranged Fe-Fe AF interactions,\cite{Kudrnovsky2015,supplement} which reduces the effective exchange parameter $J_0^{\rm Fe}$.

    However, the bct structure with $c/a$\,=\,0.843 still has a considerable $T_{\rm C}^{\rm MFA}$, which stands in
  contrast to the observed breakdown of magnetic order in the epitaxial films, but is in full agreement with the preference  for FM order in the unrelaxed super-cell calculations.    
     Introducing the orthorhombic distortions 
      enhances the competition between the different contributions, which eventually results in an effective annihilation of $J_0^{\rm Fe}$. In turn $J_0^{\rm Rh}$ finally becomes the leading contribution to $T_{\rm C}^{\rm MFA}$. 
  Observing that in the PM configuration, the $\upmu_{\rm Rh}$ collapses, while the PM Fe  moments remain well localized and decrease
  by less than $0.1\,\mu_{\rm B}$ with respect to the FM, we  conclude on a complete disappearance of FM order. 
  This is reflected by a PM ground state in the KKR-CPA calculations, confirming the super-cell results as well as the experimental observations.

   The disappearance of magnetic order   becomes more apparent from a comparison of the
    electronic DOS in the PM (Fig.~\ref{figure9})
    and the FM case (Fig.~\ref{figure10}).
    Of particular interest in the latter case is the minority channel, where we
    observe a Rh-peak slightly above the Fermi level in the cubic A2 structure.
    With increasing tetragonal distortion, this peak eventually crosses the Fermi energy
    and is finally located  slightly below $E_{\rm F}$ in the tetragonal structure
    with $c/a$$\,=\,$$0.843$. The orthorhombic distortion completely
    removes this Rh peak in the minority-spin DOS 
    -- but simultaneously another peak appears in the partial DOS of Fe close to $E_{\rm F}$. Since this feature is absent in the PM phase, it likely contributes to
    destabilization of the  FM order in the final structure, leading to the observed PM
    ground state.

In both magnetic states,
the martensitic transformation, from bct to $Cmcm$,  is accompanied by a gain in band energy, which arises from the redistribution of a peak
in the Rh partial DOS just below $E_{\rm F}$
(features marked by arrows in both magnetic configurations in Fig.~\ref{figure9} and  Fig.~\ref{figure10}).
This reminds of a band-Jahn-Teller-like mechanism, which is 
known from other ferrous  alloys, such as Fe-Pd.\cite{Opahle2009,Gruner2011}

The combination of a specific shuffling of atomic planes with the disappearance of a peak-like structure in the
minority spin DOS is reminiscent  of the Fermi surface nesting which results in a soft phonon mode via electron-phonon coupling.
This plays a decisive role in triggering the martensitic transformation in other functional
transition metal alloys and compounds, such as in the
FM shape memory alloy Ni$_2$MnGa. \cite{Lee2002a,Bungaro2003,Entel2008,Opeil2008,Haynes2012}
Particular cross-sections of the minority spin-channel of the Bloch spectral function $A(\vec{k},E_{\rm F})$ in FM FeRh (Fig.~\ref{figure11})
suggest the suppression of marked parallel features by the distortion
(a similar situation is also encountered in the PM case\cite{supplement}). 
This may be counted as a first indication of nesting-induced electron-phonon coupling.
However, to obtain a complete picture one must
evaluate the entire four-dimensional object $A(\vec{k},E_{\rm F})$
in terms of the generalized susceptibility and investigate the phonon dispersion of this disordered system, which is beyond the scope of the present work.

\section{Summary}\label{sec:sum}

In conclusion, chemically disordered FeRh was deposited epitaxially on W(001). 
A combined experimental and theoretical study indicates that the 
strain, imposed on the system via the epitaxial growth, triggers a martensitic transformation into an orthorhombic adaptive martensite.
The structural changes then lead to the total breakdown of magnetic order.
  Based on the results of our electronic structure calculations and the absence of any stable magnetic order, observed experimentally by CEMS and SQUID magnetometry,
  we propose, that the system is characterized by dynamically fluctuating disorder
  in the localized $\upmu_{\rm Fe}$ according to frustrated magnetic exchange couplings.
The frustrated magnetic exchange interactions are  a direct consequence of our epitaxial-strain tailoring of the martensitic instability.
The huge sensitivity of magnetic order to rather moderate structural changes, carefully tailored  by epitaxial strain, makes disordered FeRh particularly interesting
 for magnetostrictive or magnetoresistive sensor or actuator applications,\cite{Wilson2007} alongside its well-investigated B2-ordered counterpart.
 
 The close  resemblance to band-Jahn-Teller-type instabilities in  MSMAs indicates that this  might be a rather general feature of transition metal alloys with both, localized $d$-moment and hybridization-induced magnetic moments, accompanied by a marked feature in the DOS.
Thus, it should be observed in a variety of other systems, which can be conveniently preselected employing electronic structure calculations, by tuning the electronic features around the Fermi surface with external stimuli such as epitaxial strain.  Amplification of a symmetry breaking by an external symmetry-lowering stimulus  may be considered as a rather general and innovative  tailoring strategy for specific functional properties. An analogous case
was found in thin films of multiferroic  BiFeO$_{3}$, where the large epitaxial strains result in the formation of reduced symmetry structures,\cite{Bea2009,Dupe2010} which might be linked to a phonon softening. \cite{Liu2012a}

\begin{acknowledgments}
  We acknowledge funding by the Deut\-sche For\-schungs\-ge\-mein\-schaft via HA1344/28-1, TRR 80 and SPP 1599.
  We thank L. Bennett, E. Dellatorre (GWU Washington D.C.), P. Entel (Duisburg-Essen) and C. K\"ubel (KIT) for discussions and U. v. H\"{o}rsten (Duisburg-Essen) for  technical assistance.

\end{acknowledgments}

%

\end{document}